# Superconductivity induced by hydrogen anion substitution in 1111-type iron arsenides

Hideo Hosono,[1,2,3*] Satoru Matsuishi[3]

[1]*Materials and Structures Laboratory, Tokyo Institute of Technology, 4259 Nagatsuta-cho, Midori-ku, Yokohama 226-8503, Japan*

[2]*Frontier Research Center, Tokyo Institute of Technology, 4259 Nagatsuta-cho, Midori-ku, Yokohama 226-8503, Japan*

[3]*Materials Research Center for Element Strategy, Tokyo Institute of Technology, 4259 Nagatsuta-cho, Midori-ku, Yokohama 226-8503, Japan*

*Corresponding author: Hideo Hosono

Materials and Structures Laboratory, Tokyo Institute of Technology

4259 Nagatsuta-cho, Midori-ku, Yokohama 226-8503, Japan

TEL +81-45-924-5359

FAX +81-45-924-5339

E-mail: hosono@msl.titech.ac.jp

**Abstract**

Hydrogen is the simplest bipolar element and its valence state can be controlled from +1 to −1. We synthesized the 1111-type iron arsenides CaFeAsH and $Ln$FeAsO$_{1-x}$H$_x$ ($Ln$ = lanthanide; $0 \leq x \leq 0.5$) with the ZrCuSiAs type structure by a high-pressure synthesis method. The position and valence state of the substituted H were determined by neutron diffraction and density functional theory calculations. The close similarity in the structural and electrical properties of CaFeAsH and CaFeAsF indicated the formation of the hydride ion (H$^-$), which is isovalent with the fluoride ion (F$^-$), in the 1111-type iron arsenides. When some of the O$^{2-}$ ions in $Ln$FeAsO are replaced by H$^-$, superconductivity is induced by electron doping to the FeAs-layer to maintain charge neutrality. Since the substitution limit of hydrogen in $Ln$FeAsO ($x \approx 0.5$) is much higher than that of fluorine ($x \approx 0.2$), the hydrogen substitution technique provides an effective pathway for high-density electron-doping, making it possible to draw the complete electronic phase diagram of $Ln$FeAsO. The $x$-$T$ diagrams of LnFeAsO$_{1-x}$H$_x$ ($Ln$ = La, Ce, Sm, Gd) have a wide superconducting (SC) region spanning the range $x$ = 0.04 to 0.4, which is far from the parent antiferromagnetic region near $x$ = 0.0. For LaFeAsO$_{1-x}$H$_x$, another SC dome region was found in the range $x$ = ~ 0.2 to ~0.5 with a maximum $T_c$ = 36 K, in addition to a conventional SC dome located at $x$ ~0.08 with maximum $T_c$ = 29 K. Density functional theory calculations performed for LaFeAsO$_{1-x}$H$_x$ indicated that the newly observed $T_c$ is correlated with the appearance of degeneration of the Fe 3$d$ bands ($d_{xy}$, $d_{yz}$ and $d_{zx}$), which is caused not only by regularization of the tetrahedral shape of FeAs$_4$ due to chemical pressure effects but also by selective band occupation with doped electrons. In this article, we review the recent progress of superconductivity in 1111-type iron (oxy)arsenides and related compounds induced by hydrogen anion substitution.

# 1. Introduction

Since the discovery of the superconductivity of LaFeAsO$_{1-x}$F$_x$ ($T_c$ = 29 K) [1], layered iron pnictides and related materials have been intensively investigated as candidates for high-$T_c$ superconductors. To date, various types of iron-based superconductors have been reported [2–9]. However, the highest reported $T_c$ of iron based superconductors, ~55 K for SmFeAsO$_{1-x}$F$_x$ [10], has remained unchanged since 2008. The iron oxy-arsenides $Ln$FeAsO ($Ln$ = lanthanide) [1,10–14], so-called 1111-type iron-arsenides, have ZrCuSiAs-type structures [15,16] composed of alternate stacks of FeAs anti-fluorite-type conducting layers and $Ln$O fluorite-type insulating layers (see **Fig. 1**). At ambient pressure (AP), the parent compounds are not superconducting (SC) and superconductivity is induced by appropriate electron doping of the FeAs-layer via replacement of divalent oxygen ions with monovalent fluorine ions (O$^{2-}$ → F$^-$ + $e^-$) [10,12–14,17]. While other electron doping techniques to induce superconductivity, such as oxygen vacancy formation (O$^{2-}$ → V$_O$ + 2$e^-$) [18–20] and transition metal (Co or Ni) substitution into the iron site [21–24], have been reported, oxygen substitution by fluorine has been the most effective and practical approach to obtain high $T_c$ in 1111-type $Ln$FeAsO materials. **Figure 2** shows the schematic electronic phase diagram of $Ln$FeAsO$_{1-x}$F$_x$ as a function of $x$ and temperature $T$. In the case of LaFeAsO$_{1-x}$F$_x$, a dome-like superconducting region around $x$ ~ 0.08 is observed adjacent to the antiferromagnetic region at about $x$ = 0.0 [25,26]. However, due to the poor solubility of fluorine ($x$ < 0.2), which arises from the stable impurity phase formation of $Ln$OF, the high-$x$ side of the SC dome is hidden and the upper critical electron-doping level for superconductivity has not yet been established.

Recently, we found an alternative technique that is capable of high concentration electron doping far beyond the limit of fluorine substitution [27–29]. Using the stability of the negatively charged state of hydrogen, i.e., H$^-$, in rare-earth compounds, the oxygen site in $Ln$FeAsO was successfully replaced by hydrogen with $x$ up to ~ 0.5. The hydrogen substitution not only follows the results of fluorine substitution at $x$ < 0.2 but it also suggested that the superconducting region continues up to $x$ ~ 0.5. In particular, the LaFeAsO$_{1-x}$H$_x$ system shows a SC dome region located at $x$ ~ 0.35 with a maximum $T_c$ = 36 K, in addition to the conventional superconducting region located at $x$ ~ 0.08. In this article, application of the hydrogen substitution technique to 1111-type arsenides is discussed by reviewing recent works on 1111-type CaFeAsH [27] and $Ln$FeAsO$_{1-x}$H$_x$ materials [27–29].

## 2. Iron arsenide hydrides with ZrCuSiAs-type structure

In mid-2008, the drastic improvement of the $T_c$ of LaFeAsO$_{1-x}$F$_x$ due to $Ln$-substitution intensified the effort to synthesize analogues of 1111-type iron arsenide with alternative blocking layers in the place of the $Ln$O-layer. Considering the formal charged state of the $Ln$O and FeAs layers ($Ln^{3+}$O$^{2-}$ and Fe$^{2+}$As$^{3-}$), we succeeded in synthesizing $Ae$FeAsF ($Ae$ = Ca, Sr), which has an $Ae$F layer with the same formal charge of +1 ($Ae^{2+}$F$^-$) as the $Ln$O layer [30,31]. Soon after our results were published, several groups reported the synthesis of SrFeAsF and EuFeAsF [32–34]. Like $Ln$FeAsO, stoichiometric $Ae$FeAsF materials are antiferromagnetic metals and superconductivity can be induced by partial replacement of Fe with Co or Ni (maximum $T_c$ = 22 K) [30,31,35–38]. These properties indicate that the FeAs layer in $Ae$FeAsF is isoelectronic to that in $Ln$FeAsO. Several groups reported the superconductivity of $Ae$FeAsF with $T_c$ of up to 56 K induced by $Ln$ substitution in the $Ae$ site [34,39,40]. However, the details of the superconductivity have not been reported, even though the $T_c$ is very high. To further extend the series of 1111-type iron arsenides, we searched for other combinations of ions that could form a blocking layer. From the research on the cultivation of electro-active functionality in anion-encaging 12CaO·7Al$_2$O$_3$ crystals [41–44], we noticed that the hydrogen anion can be stabilized in the oxides if electropositive cations coordinate to hydrogen and no other stable anion is supplied. In fact, the hydride anion is stable in the compounds of electropositive elements such as alkali, alkali-earth and rare-earth metals. We noticed that the blocking layers of 1111-type compounds are also composed of these elements. Thus, the synthesis of prototypical 1111-type arsenide "hydrides" $Ae$FeAsH with formal charges of $Ae^{2+}$Fe$^{2+}$As$^{3-}$H$^-$ was attempted.

Some 1111-type hydrides, $LnMX$H ($M$ = Mn, Fe, Co, Ru; $X$ = Si, Ge).[45–51], have already been reported These compounds have been synthesized by the insertion of hydrogen into 111-type $LnMX$ with the CeFeSi-type structure [52], which are composed of $Ln$ mono-atomic layers and $MX$ layers that are isostructural to the FeAs layer in $Ln$FeAsO. By heating in high-pressure hydrogen gas, $LnMX$ is converted to $LnMX$H. Using this approach, we succeeded in synthesizing $Ln$-free 1111-type CaNiGeH by the insertion of hydrogen into 111-type CaNiGe [53]. However, this technique is not applicable to the synthesis of 1111-type $Ae$FeAsH because 111-type $Ae$FeAs compounds are not possible.

We developed the high-pressure technique using metal hydride as a component of the starting mixtures [27]. 1111-type CaFeAsH and its solid solution with CaFeAsF (CaFeAsF$_{1-x}$H$_x$) were directly synthesized by the solid-state reaction of metal arsenides (CaAs, Fe$_2$As) and hydrides (CaH$_2$) using a belt-type high-pressure anvil cell: CaAs + Fe$_2$As + CaH$_2$ → 2 CaFeAsH. CaAs and Fe$_2$As were synthesized by the reaction of the respective metals with arsenic, and $CaH_2$ was synthesized by heating Ca metal in hydrogen gas. The starting mixture was prepared in a glove box filled with purified Ar gas ($H_2O$, $O_2$ < 1 ppm) and placed in a BN capsule with $LiAlH_4$ as an excess solid hydrogen source. By heating the mixture at 1273 K and 2 GPa for 30 min, tetragonal 1111-type CaFeAsH was successfully obtained in a polycrystalline pellet form. The deuterated analog CaFeAsD was also synthesized using $CaD_2$ and $LiAlD_4$, and the differences in the lattice parameters of the resulting hydride and deuteride samples were less than 0.05%. Since CaFeAsH is decomposed into $CaFe_2As_2$, Ca and $H_2$ when heating above 573 K, the hydrogen content in the resulting compounds were estimated by thermogravimetry/mass spectroscopy (TG/MS). As shown in **Fig. 3**, weight loss due to the emission of $H_2$ molecules ($m/z$ = 2) was observed from 573 K to 873 K. The amount of $H_2$ released was estimated to be 3.08 mmol/g from the integration of the mass peak of $m/z$ = 2. This quantity is almost equal to that expected for the decomposition of CaFeAsH within the measurement error range (2CaFeAsH → Ca + $CaFe_2As_2$ + $H_2$, 2.91 mmol/g). **Table 1** shows the crystallographic parameters of CaFeAsD determined by Rietveld refinement of the neutron powder diffraction patterns observed at 300 K and 10 K. The anion site in the block layer is occupied by deuterium with a site occupancy of 0.935. Taking into account the isotopic purity of $LiAlD_4$ and the inclusion of hydrogen from other starting materials, we concluded that the remaining fraction (0.065) of the anion sites is mainly occupied by proton nuclei. At 10 K, an orthorhombic phase with space group *Cmma* was determined. The change of crystal symmetry from tetragonal to orthorhombic is typically observed in the 1111 type iron arsenides *Ln*FeAsO and *Ae*FeAsF.

The CaFeAsH obtained was semi-metallic with electrical conductivity of ~2.5 mΩ·cm at 300 K. **Figure 4** shows the temperature dependence of dc electrical resistivity (ρ) of CaFeAsH normalized by the room temperature resistivity ($ρ_{300K}$). $ρ/ρ_{300K}$ of CaFeAsF is also plotted for comparison. The ρ-*T* profile of CaFeAsH is very similar to that of CaFeAsF and characterized by the sudden decrease of resistivity at $T_s$ = 120 K, corresponding to the tetragonal to orthorhombic structural transition associated with antiferromagnetic transition with Néel temperature $T_N$ = 113 K [36,37]. This similarity strongly suggests that the structural and magnetic transitions also occur in CaFeAsH.

**Figure 5** shows the total and projected density of states (DOS and PDOS) of CaFeAsH obtained from density functional theory (DFT) calculations. The origin of the horizontal axis is set at the Fermi level ($E_F$). By minimizing the total energy with respect to both the coordinates of all atoms and the lattice parameters, the stripe-type antiferromagnetic ordering of the Fe spins, commonly observed for 1111-type iron arsenide, was obtained as the most stable magnetic structure. The total DOS and Fe PDOS profiles are consistent with the semi-metallic nature of CaFeAsH. Hydrogen 1s states were located ~2 eV below the $E_F$ and were fully occupied. It is evident from these results that hydrogen is present as $H^-$ ($1s^2$) in CaFeAsH. This finding is consistent with the similarities in the ρ-$T$ curves of CaFeAsH and CaFeAsF, indicating that the replacement of F sites with $H^-$ ions does not significantly affect the electronic structure of the FeAs conduction layer.

**Figure 6** shows the lattice parameters $a$ and $c$ of the solid solution of CaFeAsF and CaFeAsH (CaFeAsF$_{1-x}$H$_x$) as a function of hydrogen content $x$. While the lattice parameter $a$ is almost independent of $x$, the value of $c$ is proportional to $x$, indicating that the geometry of the CaF$_{1-x}$H$_x$ layer is determined by the weighted average of the ionic radii of $F^-$ ($r_F$) and $H^-$ ($r_H$). Assuming that $F^-$ coordinated by four Ca$^{2+}$ ions retains Shannon's ionic radius ($r_F$ = 131 pm), the ionic radius of Ca$^{2+}$ calculated from the F-Ca distance in CaFeAsF ($r_F + r_{Ca}$ = 234 pm) is 103 pm. The radius of $H^-$ then calculated from the H-Ca bond length (230 pm) in CaFeAsH is 127 pm, which is 3% smaller than that of $F^-$. The above results demonstrate the existence of 1111-type arsenide hydrides and the high similarity of hydride and fluoride ions in 1111-type compounds. While superconductivity of CaFeAsH has not yet been observed, appropriate electron-doping methods, such as transition metal substitution, will make the material superconductive.

## 3. Hydrogen substitution into LnFeAsO

The similarity of fluorine and hydrogen in CaFeAsF$_{1-x}$H$_x$ suggests that hydrogen could be an alternative dopant anion to induce superconductivity of $Ln$FeAsO. Like CaFeAsH, hydrogen-substituted $Ln$FeAsO ($Ln$FeAsO$_{1-x}$H$_x$) was synthesized by the solid-state reaction of metal arsenides ($Ln$As, Fe$_2$As, and FeAs), oxides ($Ln_2$O$_3$ and CeO$_2$) and hydrides ($Ln$H$_2$) under high-pressure and temperature (2 GPa, 1473 K). $Ln$H$_2$ was synthesized by heating $Ln$ metal powder in hydrogen gas under an ambient atmosphere.

**Figure** 7(a) shows the powder XRD patterns of SmFeAsO$_{1-x}$H$_x$ synthesized with nominal $x$ ($x_{nom}$) of 0.3 and 0.4. In the case of fluorine substitution, the segregation of $Ln$OF prevents the formation of 1111-type $Ln$FeAsO$_{1-x}$F$_x$ with $x >$ ~0.15. In contrast, for SmFeAsO$_{1-x}$H$_x$, the segregation of the impurity phase (SmAs) was only confirmed at $x_{nom} > 0.4$. The hydrogen in $Ln$FeAsO$_{1-x}$H$_x$ desorbs in the form of hydrogen molecules upon heating at >600 K, and the amount can be measured by TG/MS (see Fig. 7(b)). Figure 7(c) compares the measured hydrogen content ($y$) and the oxygen deficiency ($x$) evaluated by electron-probe microanalysis (EPMA) for the chemical formula SmFeAsO$_{1-x}$H$_y$. With increasing $x_{nom}$ in the starting mixture, both $x$ and $y$ increased in the relationship $y \sim x$. This relationship is commonly observed for hydrogen-substituted $Ln$FeAsO and is independent of $Ln$. **Figure 8** shows the variation of the lattice parameters $a$ and $c$ in $Ln$FeAsO$_{1-x}$H$_x$ ($Ln$ =La, Ce, Sm, Gd) as a function of $x$ measured by EPMA. Like CaFeAsF$_{1-x}$H$_x$, the $c$ parameter length linearly decreases with increasing $x$, indicating that the O$^{2-}$ is successfully replaced by H$^-$. While the lattice parameter $a$ of CaFeAsF$_{1-x}$H$_x$ was independent of $x$, for $Ln$FeAsO$_{1-x}$H$_x$ it significantly decreases with increasing $x$. This indicates that hydrogen substitution into $Ln$FeAsO changes the bonding state of the FeAs layer, i.e., electrons are donated to the FeAs-layer (O$^{2-}$ → H$^-$ +e$^-$), while H$^-$ substitution into CaFeAsF (F$^-$ → H$^-$) does not involve electron donation to the FeAs layer.

The replacement of oxygen by hydrogen has been experimentally confirmed by neutron powder diffraction. The diffraction patterns of CeFeAsO$_{1-x}$D$_y$ with $x$ = 0.15, 0.28 and 0.43 estimated by EPMA are in good agreement with the simulated patterns of structural models in which deuterium nuclei occupy the oxygen sites with occupancy $y$ = 0.13, 0.24 and 0.37, respectively. Here, we noticed that the impurity proton was contained in the excess deuterium source (LiAlH$_4$) for high-pressure synthesis ([D]: [H] ~ 9: 1), and the underestimation of $y$ compared with $x$ (−14 %) can be explained by the effect of impurity protons. Therefore, the above results indicate that the occupancy of hydrogen species is close to $x$, and the oxygen vacancies are filled by hydrogen to form $Ln$FeAsO$_{1-x}$H$_x$.

To confirm the valence state of hydrogen in the oxygen sites, we performed DFT calculations. **Figure 9**

shows the calculated DOS of CeFeAsO, CeFeAsO$_{0.75}$F$_{0.25}$ and CeFeAsO$_{0.75}$H$_{0.25}$ assuming the stripe-type antiferromagnetic ordering of Fe spins and the checkerboard-like ordering of Ce spins. For CeFeAsO$_{0.75}$F$_{0.25}$ and CeFeAsO$_{0.75}$H$_{0.25}$, 25% of the oxygen sites are replaced by fluorine or hydrogen. The lattice parameters, atomic positions and local spin moments were fully relaxed by a structural optimization procedure that minimizes the total energy and forces on the atoms. In contrast to the experimental results, magnetic ordering remained in CeFeAsO$_{0.75}$F$_{0.25}$ and CeFeAsO$_{0.75}$H$_{0.25}$. This discrepancy is attributed to due to DFT using conventional functionals being inadequate for reproducing the low energy phenomena in the Fe-arsenide system. However, the calculations are still useful for determining the bonding states. In CeFeAsO$_{0.75}$F$_{0.25}$, fluorine forms a narrow isolated 2$p$ band located −8 eV from the $E_F$, while the hydrogen in CeFeAsO$_{0.75}$H$_{0.25}$ forms a broad 1$s$ band located between −3 and 6 eV from the $E_F$, which is close to that of the oxygen 2$p$ band and indicates that the energies of the hydrogen 1$s$ and oxygen 2$p$ orbitals overlap and form a unified valence band. The integrated PDOS for the hydrogen 1s band is 0.5 electrons per f.u., corresponding to the H$^−$ state (1$s^2$). Furthermore, fluorine or hydrogen substitution increases the $E_F$ by 0.3 eV, corresponding to a doping of 0.25 electrons per Fe. We therefore conclude that hydrogen substituted into the oxygen site forms H$^−$, and an electron is donated to the FeAs layer, as with F$^−$ substitution.

**Figure 10**(a) shows the ρ-$T$ curves of SmFeAsO$_{1−x}$H$_x$. A sudden decrease in ρ due to the superconducting transition is observed for $x \geq 0.03$, and the maximum $T_c$ of 55 K is recorded for $x \sim 0.2$. When $x$ is increased above 0.2, $T_c$ decreases, indicating the appearance of an over-doped region. To clarify the change in transport properties of the normal conducting state with hydrogen substitution, we examined the exponent $n$ of ρ($T$) = ρ$_0$ + $T^n$ in the temperature range from just above $T_c$ and below 130 K. With increasing $x$ below the optimal substitution level ($x < 0.2$), $n$ decreases from 2, which is characteristic of strongly correlated Fermi liquid systems, to 1, corresponding to the non-Fermi liquid state. Similar behavior has been observed in 122- and some 1111-type iron arsenides [54,55]. In contrast, $n$ deceases to 0.5 with increasing $x$ in the over-doped region ($x > 0.2$). This is quite different from 122-type arsenides where $n$ increases to ~2 in the over-doped region.

Figure 10(b) shows the electron doping level $x$ vs. $T$ phase diagrams of SmFeAsO$_{1−x}$H$_x$ and SmFeAsO$_{1−x}$F$_x$ [56]. The $T_c$ vs. $x$ curves of SmFeAsO$_{1−x}$H$_x$ and SmFeAsO$_{1−x}$F$_x$ overlap at $x < 0.15$, indicating that hydrogen substitution results in indirect electron doping of the FeAs layer as with fluorine substitution. While the solubility limit of fluorine in the oxygen site is restricted to less than 20 % ($x = 0.2$), that of hydrogen can reach ~40 %. This wider substitution range makes it possible to optimize the electron doping level to induce superconductivity, and to complete the electronic phase diagram including the over-doped region.

## 4. Two SC domes amd Electronic structure of LaFeAsO$_{1-x}$H$_x$

**Figure 11** shows the wide-range electronic phase diagram of electron-doped $Ln$FeAsO with $Ln$ = La, Ce, Sm and Gd determined by high concentration doping via hydrogen substitution. As references, $T_c$ and $T_s$ of $Ln$FeAsO$_{1-x}$F$_x$ are plotted on the diagrams. From the results of fluorine substitution, the SC region forms a simple dome-like shape adjacent to the AFM in the parent phase. This result is consistent with the spin-fluctuation-mediated mechanism of superconductivity proposed for the FeAs-based superconductor [57,58]. However, our results for hydrogen-substitution reveal that the SC region extends above $x = 0.3$ and far away from the parent AFM region. Furthermore, the SC region for LaFeAsO$_{1-x}$H$_x$ forms a double dome shape composed of the conventional narrow SC dome located around $x = 0.08$ with a maximum $T_c = 29$ K and an additional wide SC dome located around $x = 0.35$ with a maximum $T_c = 36$ K. To characterize the nature of this two-dome structure, we examined the exponent $n$ of $\rho(T) = \rho_0 + T^n$ for the first and the second domes. **Figure 12** compares the $\rho$-$T$ curves for $x = 0.08$ and 0.35. In contrast to the normal conducting state of $x = 0.08$ that shows $T^2$-dependence of $\rho$, $x = 0.35$ shows $T^1$-dependence, which is the same as the optimally electron-doped state of $Ln$FeAsO$_{1-x}$H$_x$ with $Ln \neq$ La. This indicates that the superconductivity observed in the second SC dome originates from the common electronic state that is typical of the SC domes in other 1111-type $Ln$FeAsO ($Ln \neq$ La). In other word, the first SC dome has exceptional features, although its electronic state has been considered as a standard or prototype to explain the mechanism of superconductivity in iron-based superconductors. When the high pressure was applied, the local minimum in $T_c$ at about $x = 0.21$ under ambient pressure disappears and the two-dome shape of the SC region joins to give a single SC dome with a maximum $T_c$ of 46 K as shown in **Fig.13**. The shape and the maximum $T_c$ value are similar to those of CeFeAsO$_{1-x}$H$_x$ at ambient pressure but with smaller lattice parameters. This joining of the two domes upon the application of high pressure can be attributed to lattice compression, with the decrease of the lattice parameter $a$ (~ 1%) in LaFeAsO$_{1-x}$F$_x$ bringing it close to a LaFeAsO$_{1-x}$H$_x$ lattice [59]. A further 1% decrease of the lattice parameter $a$ of LaFeAsO$_{1-x}$H$_x$ brings its lattice close to that of CeFeAsO$_{1-x}$H$_x$ under ambient pressure. This means that the shape of the SC region is related to the local coordination structure, such as the symmetry of the FeAs$_4$ tetrahedron.

To investigate the electronic structure of the highly electron doped state of LaFeAsO$_{1-x}$H$_x$ with $x > 0.2$, we performed DFT calculations using the WIEN2K code22 employing the generalized gradient approximation Perdew–Burke–Ernzerhof functional23 and the full-potential linearized augmented plane wave plus localized orbitals method. To ensure convergence, the linearized augmented plane wave basis set was defined by the cutoff RMTKMAX = 9.0 (RMT: the smallest atomic sphere radius in the unit cell), with a mesh sampling of 15×15×9 k points in the Brillouin zone. For H-substituted samples we employed virtual crystal approximation (VCA) where hydrogen behaves like fluorine and increases the nuclear charge of the oxygen site $x$. Using VCA, the effect of electron doping can be self-consistently computed, where not only the shift of the $E_F$ but also the change of the band dispersion are simulated. **Figure 14**(a) shows 2-dimensional cross sections of FS for doping levels corresponding to the top of the first SC dome ($x = 0.08$), the $T_c$ valley (0.21), the top of the second SC dome (0.36), and the over-doped region (0.40). In the calculations, the lattice parameters and atomic positions are fixed at the experimental values determined by XRD at 20 K. At $x = 0.08$, the size of an outer $d_{xy}$ ($x$, $y$, and $z$ coordination is given by the Fe square lattice) or inner $d_{yz,zx}$ hole pocket at the Γ point is close to that of the two electron pockets at the M point, indicating that the nesting between the hole and the electron pockets in the (π, π) direction is strong. The spin-fluctuation-mediated mechanism is based on the FS nesting [57,58]. As $x$ increases, the nesting is monotonically weakened because the hole pockets gradually contract while the electron pockets expand. It has been observed that as the pnictogen height, $h_{Pn}$, from the Fe plane increases, the $d_{xy}$ hole pocket expands, and the nesting condition becomes better [60,61]. In the present case, although $h_{As}$ increases with $x$ as shown in **Fig. 15** (b), the size of the $d_{xy}$ hole pocket remains almost unchanged irrespective of $x$. This result can be explained by considering that expansion of the $d_{xy}$ hole pocket due to structural modification is canceled out by contraction due to the up-shift of $E_F$ by electron doping to this band.

Nesting between the hole and electron pockets is the most important attraction in the spin fluctuation mechanism. The decrease in $T_c$ when $x$ is increased from 0.08 to 0.21 is the result of reduction in the spin fluctuations due to weakening of the nesting in a similar way to LaFeAsO$_{1-x}$F$_x$. However, the present spin fluctuation mechanism cannot explain the experimental findings that $T_c$ ($x$) also increases in the range 0.21 < $x$ < 0.53 and has a maximum of 36 K at $x \sim 0.36$.

Figure 14 (b) shows the band structures around $E_F$ for $x$ = 0.08, 0.21, 0.36, and 0.40. Since the unit cell contains two Fe atoms, there are 10 bands around the $E_F$ due to bonding and anti-bonding of the 3$d$ orbitals of the two neighboring Fe atoms, which cross the $E_F$ around the Γ and M points. Figure 15(a) shows the energy shift of specific bands at various Γ points (Γ$_{anti-dxy}$, Γ$_{dxy}$ and Γ$_{dyz,zx}$) as a function of $x$. The unoccupied bands continuously decrease in energy with increasing $x$. In particular, the bands derived from the anti-bonding orbital between the $d_{xy}$ orbitals, which hereafter we will call the "anti-$d_{xy}$ bands", and the degenerate $d_{yz,zx}$ bonding orbitals decrease in energy and cross the bonding-$d_{xy}$ band at about $x$ = 0.36, forming a triply degenerated state near $E_F$. After the degenerate state is formed, these anti-$d_{xy}$ and bonding $d_{yz,zx}$ bands create a new band below $E_F$ at $x$ = 0.40. Note that, in Fig. 15(b) the As-Fe-As angle of the FeAs$_4$ tetrahedron is significantly deviated from the ideal tetrahedral angle (109.3˚) in the optimally doped region ($x$ = 0.33–0.46). Using a rigid band model without VCA, the band-crossing is not well explained, indicating that the energy shift of the relevant Fe 3$d$ bands is not determined by only the change in the local structure around iron but also by the asymmetric occupation of doped electrons in the bonding-$d_{yz, zx}$, $d_{xy}$ and anti-bonding-$d_{xy}$ orbitals. Since the bonding $d_{yz,zx}$ and $d_{xy}$ bands are almost flat in the Γ-Z direction, the band crossing at $x$ = 0.36 forms a shoulder in the total density of states (DOS) at $E_F$ (Fig. 14(c)), indicating the presence of electronic instability arising from degeneration of the $d_{xy}$ and $d_{yz,zx}$ orbitals. In this situation, structural transitions, such as band Jahn-Teller distortions, can occur and decrease the energy of the system. However, structural transitions were not observed by XRD using a Synchrotron X-rays source for temperatures >20 K in samples with 0.08 ≤ $x$ ≤ 0.40.

Considering these results, the band degeneracy appears to have an important role in the emergence of the second dome. For the iron-based superconductors, orbital fluctuation is a plausible alternative pairing model, which was derived from the large softening of the shear modulus observed near the tetragonal–orthorhombic transition of parent compounds. If we follow the orbital fluctuation model, the second dome and $T$-linear resistivity in the present system might be understood as the result of electron pairing and carrier scattering by the fluctuations of the degenerated Fe-3$d_{xy,yz,zx}$ orbitals, respectively. The discovery of a two dome structure indicates that improvement of the theory of the superconducting mechanism for Fe-based superconductor is required.

## 5. Hydrogen-substitution in other 1111-type arsenides

In addition to 1111 type iron-arsenides, we also applied the hydrogen-substitution technique to other 1111-type transition metal arsenides. 1111-type LaNiAsO is a superconductor with $T_c$ = 2.4 K, and the $T_c$ can be increase to 3.7 K by fluorine substitution in the oxygen sites [62,63]. However, the substitution limit of fluorine in LaNiAsO$_{1-x}$F$_x$ is less than 15%, and the complete picture of the superconducting dome in the electronic phase diagram has not been determined. Recently, LaNiAsO$_{1-x}$H$_x$ was successfully synthesized by the solid state reaction of LaAs, NiAs, Ni$_2$As, La$_2$O$_3$, and LaH$_2$ at 1273-1373 K and 2 GPa [64]. Up to 18% of the oxygen sites were substituted by hydrogen and the lattice parameters $a$ and $c$ monotonically decreased with increasing $x$. $T_c$ could be increased to 3.7 K with $x$ = 0.15, but further increase of $x$ decreased $T_c$ to 1.7 K due to over-doping (see **Fig. 16**). While the $T_c$ of LaNiAsO can be decreased by applying a physical or chemical pressure due to isovalent substitution of La with other $Ln$ (Ce, Pr or Nd), an increase of $T_c$ is only observed in fluorine or hydrogen substituted systems. This indicates that hydrogen and fluorine substitution results in electron doping of the NiAs layer via the formation of monovalent ions of H$^-$ and F$^-$ at the O$^{2-}$ site.

Recently, we applied the hydrogen-substitution technique to LaMnAsO [65]. $Ln$MnAsO is an antiferromagnetic insulator/semiconductor with an optical band gap of ~1 eV and a checkerboard-type ordering of Mn spin below $T_N$ = 300 K [66–68]. In past studies, electron doping via fluorine substitution in oxygen sites to form $Ln$MnAsO$_{1-x}$F$_x$ and the creation of oxygen vacancies to form $Ln$MnAsO$_{1-x}$ have been performed to modify the electronic properties [69,70]. The reduction of the electrical resistivity accompanied with colossal magnetoresistance (CMR) was induced by carrier doping. However, the antiferromagnetic ordering has not been suppressed, indicating that the maximum concentration of doped electrons is small ($x$ < ~0.1). In contrast to conventional techniques, the substitution limit of hydrogen is as high as $x$ = 0.73 and the antiferromagnetic ordering is suppressed by hydrogen substitution. For $x$ > 0.08, the insulating/semiconducting state was converted into a metallic state, and ferromagnetism was induced by the direct exchange interaction between neighboring manganese atoms mediated by doped electrons. With increasing $x$, the Curie temperature $T_C$ increased up to 264 K at $x$ = 0.73, and the local moment per Mn atom reached 1.5 $\mu_B$. Large negative magnetoresistance up to 60% was observed at the boundary between the ferromagnetic and paramagnetic metal regions as with CMR LaMnO$_3$ [71]. The above results indicate the potential ability of the hydrogen substitution technique for modifying the electronic properties of compounds with the ZrCuSiAs-type structure [15]. Recently, Kobayashi et al. reported the synthesis of a perovskite-type BaTiO$_{3-x}$ ($x \leq 0.6$) as an O$^{2-}$/H$^-$ solid solution [72]. Up to 20 % of the oxygen ions in the typical oxide can be substituted by hydride ions. This indicates the beginning of novel solid state chemistry and physics of oxide-based compounds by substituting $O^{2-}$ with $H^-$.

## 6. Summary and Perspective

Based on material design using the stability of the hydride ion, we succeeded in synthesizing the hydrogen containing 1111-type iron arsenides CaFeAsF$_{1-x}$H$_x$, $Ln$FeAsO$_{1-x}$H$_x$, $Ln$NiAsO$_{1-x}$H$_x$ and LaMnAsO$_{1-x}$H$_x$. The high substitution limit and affinity of hydrogen for the oxide block layer gives us a route for heavy electron doping of 1111-type compounds. For $Ln$FeAsO and $Ln$NiAsO, the complete SC region in the electronic phase diagrams due to doped electrons was determined by hydrogen substitution in the oxygen sites. The unexpected shape and width of the SC region of LaFeAsO$_{1-x}$H$_x$ suggests that revision of the superconducting mechanism proposed for Fe-based superconductors is required. For LaMnAsO, suppression of the antiferromagnetism accompanied with ferromagnetic transition and insulator to metal conversion was observed. It indicates that the hydrogen substitution technique has the ability to clarify the properties of 1111-type compounds, which is not possible in the stoichiometric state.

It is intriguing to note that hydrogen anion may substitute oxygen ion site in oxides. Although we described the extensive substitution of O$^{2-}$ sites of 3d transition metal oxynpnictides with H- applying a high pressure technique, this substitution appears to occur in most oxides. For example, ZnO is a representative N-type semiconductive oxide semiconductor the possibility of electron donor but the origin of electron donor remains unclear yet in ZnO. Van der Waale [73] suggested that impurity hydrogen works donor by theoretical calculations and this suggestion was compatible with muon experiments.[74] Hydrogen is an abundant impurity but its experimental identification is rather difficult. As described in this article, it is expected that thermal desorption technique combined with DFT calculations would be helpful for quantification and charge state of incorporated hydrogen. We think that a hydrogen anion is a hidden electron donor in many oxide materials. This idea, of course, indicates the beginning of novel solid state chemistry and physics of oxide-based compounds by substitution O$^{2-}$ with H$^-$

**Acknowledgments**

The authors thank T. Hanna, Y. Muraba and S. Iimura of our laboratory for their cooperative efforts. This research was funded by the Japan Society for the Promotion of Science through the FIRST program, initiated by the Council for Science and Technology Policy, Japan. A part of this study was also supported by the Element Strategy Initiative Project for Research Core Formation, MEXT, JAPAN.

**Figure Captions**

**Figure 1.** Crystal structure of 1111-type $Ln$FeAsO ($Ln$ = lanthanide). (a) FeAs conducting layer composed of edge-shared FeAs$_4$ tetrahedra sandwiched between $Ln$O insulating layers composed of edge-shared La$_4$O tetrahedra. By replacing the $Ln$O layer with $Ae$F ($Ae$ = alkali-earth metal) or CaH layers, 1111-type $Ae$FeAsF and CaFeAsH, analogs of $Ln$FeAsO, are synthesized. (b) Fe-square lattice in FeAs-layer. In the stoichiometric compounds, the stripe-type antiferromagnetic ordering with an orthorhombic lattice is observed below 110-160 K.

**Figure 2.** Typical electronic phase diagram of $Ln$FeAsO$_{1-x}$F$_x$. AFM, PM and SC indicate the antoferromagnetic, paramagnetic and superconducting regions, respectively. Fluorine substitution to oxygen sites results in the donation of excess electrons to the FeAs-layer (O$^{2-}$ → F$^-$ + $e^-$). With increasing $x$, the tetragonal to orthorhombic structural transition with transition temperature $T_s$ and the antiferromagnetic transition with Néel temperature $T_N$ are suppressed, and superconductivity is then induced below the critical temperature of $T_c$. Since the substitution limit of fluorine is restricted when $x$ < ~0.2 due to the segregation of the $Ln$OF phase, the heavily electron doped side of the superconducting region has not been investigated.

**Figure 3.** TG and MS ($m/z$ = 2 corresponds to the H$_2$ molecule) profiles of CaFeAsH. Weight loss due to hydrogen desorption arising from sample decomposition was observed from 200 to 600 °C. The hydrogen concentration was estimated to be 1.05 molecules per unit cell by integration of the MS curve.

**Figure 4.** ρ-$T$ profile of CaFeAsH compared with CaFeAsF.

**Figure 5.** Total and atomic projected density of states of CaFeAsH obtained by density functional theory. The origin was set to the Fermi level ($E_F$).

**Figure 6.** Variation of the lattice parameters $a$ and $c$ with $x$ in the solid solution CaFeAsF$_{1-x}$H$_x$. The hydrogen content ($x$) was determined by the TG/MS method.

**Figure 7.** Structural and chemical composition data of SmFeAsO$_{1-x}$H$_x$. (a) Powder XRD patterns of SmFeAsO$_{1-x}$H$_x$ (nominal $x$ = 0.3, 0.4 and 0.5). The insets show enlarged views. (b) TG and MS ($m/z$ = 2)

profiles with nominal $x$ = 0.15. $H_2$ gas emission associated with weight loss was observed from 400 to 800 °C. (c) Oxygen deficiency content ($x$) determined by EPMA measurement and hydrogen content ($y$) estimated by the TG-MS method of SmFeAsO$_{1-x}$H$_y$ as a function of nominal $x$ ($x_{nom}$) in the starting mixture. The measured $y$ is almost equal to $x$ and $x_{nom}$, indicating that the vacancies of the oxygen sites are fully occupied by hydrogen.

**Figure 8.** Lattice parameters $a$ and $c$ of $Ln$FeAsO$_{1-x}$H$_x$ ($Ln$ = La, Ce, Sm, and Gd) as a function of $x$.

**Figure 9.** Total DOS and atomic projected DOS (PDOS) of CeFeAsO, CeFeAsO$_{0.75}$F$_{0.25}$ and CeFeAsO$_{0.75}$H$_{0.25}$ obtained by DFT calculations. The energy origin is set at the Fermi level ($E_F$).

**Figure 10.** Temperature dependence of the electrical resistivity and the electronic phase diagram of SmFeAsO$_{1-x}$H$_x$. (a) ρ-$T$ curves of SmFeAsO$_{1-x}$H$_x$ in under-doped (left; $x$ = 0.0-0.19) and over-doped states (right; $x$ = 0.22-0.47). (b) $x$-$T$ diagrams of SmFeAsO$_{1-x}$H$_x$ and SmFeAsO$_{1-x}$F$_x$.

**Figure 11.** Superconducting transition temperature $T_c$ and ρ-$T$ anomaly temperature $T_s$ of $Ln$FeAsO$_{1-x}$H$_x$ ($Ln$ = La, Ce, Sm, and Gd) as a function of $x$. $T_c$ and $T_s$ of $Ln$FeAsO$_{1-x}$F$_x$ are also plotted. An arrow denotes the optimal Tc for each system.

**Figure 12.** ρ-$T$ curves of LaFeAsO$_{1-x}$H$_x$ with $x$ = 0.08 and $x$ = 0.35. The inset shows the ρ-$T$ curve of SmFeAsO$_{0.83}$H$_{0.17}$.

**Figure 13.** Effect of pressure on the $T_c(x)$ of LaFeAsO$_{1-x}$H$_x$. ■：data obtained at ambient pressure, ▽：data at 3GPa.

**Figure 14.** Electronic structure of LaFeAsO$_{1-x}$H$_x$. (a) 2-dimensional Fermi surface of LaFeAsO$_{1-x}$H$_x$ with $x$ = 0.08, 0.21, 0.36, and 0.40. The arrow represents the nesting vector in the (π, π) direction. The contribution of the Fe-$d_{xy}$ and $d_{yz,zx}$ orbitals are colored blue and red, respectively (Ref. 28). (b) Band structures of LaFeAsO$_{1-x}$H$_x$ with $x$ = 0.08, 0.21, 0.36 and 0.40. (c) Total DOS and partial DOS of the $d_{xy}$ and $d_{yz,zx}$ orbitals.

**Figure 15.** (a) Variation in the energy level of relevant Fe 3$d$ bands at the $\Gamma$ point with $x$. The inset is the band structure of LaFeAsO$_{0.92}$H$_{0.08}$. Each $d_{xy}$ denotes an energy state for a bond composed of two Fe $d_{xy}$ orbitals in a unit cell forming an anti-bonding level. (b) Hydrogen doping dependence of the As-Fe-As angle in the FeAs$_4$ tetrahedron and arsenic height $h_{As}$ from the Fe plane. The angles and $h_{As}$ are determined from synchrotron X-ray diffraction measurements at 20 K.

**Figure 16.** $T_c$ vs. $x$ plots of LaNiAsO$_{1-x}$H$_x$ synthesized at 2GPa and LaNiAsO$_{1-x}$F$_x$ synthesized at ambient pressure.

**Table 1.** Crystallographic parameters of CaFeAsD at (a) 300 K and (b) 10 K.

(a) 300 K

$P4/nmm$, $a = 0.38763$ (1), $c = 0.82549$ (2) nm

| Atom | x   | y   | z         | g        | $B_{iso}$ (A) |
|------|-----|-----|-----------|----------|---------------|
| Ca   | ¼   | ¼   | 0.1481(4) | 1        | 1.10(7)       |
| Fe   | ¾   | ¼   | ½         | 1        | 0.81(4)       |
| As   | ¼   | ¼   | 0.6719(3) | 1        | 0.59(4)       |
| D    | ¾   | ¼   | 0         | 0.935(5) | 1.37(6)       |

(b) 10 K

$C$mma, $a = 0.54921(2)$, $b = 0.54566$ (2),

$c = 0.82115$ (2) nm

| Atom | x   | y   | z         | g        | $B_{iso}$ (A) |
|------|-----|-----|-----------|----------|---------------|
| Ca   | 0   | ¼   | 0.1472(4) | 1        | 0.83(7)       |
| Fe   | ¼   | 0   | ½         | 1        | 0.38(4)       |
| As   | 0   | ¼   | 0.6718(3) | 1        | 0.16(4)       |
| H    | ¼   | 0   | 0         | 0.918(6) | 1.16(7)       |

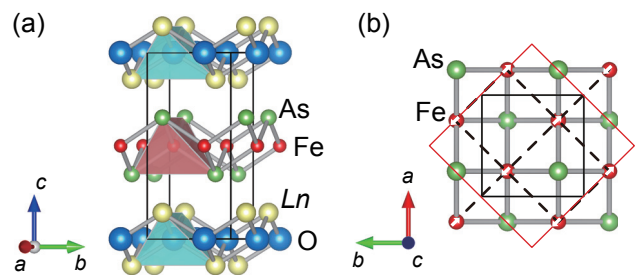

Fig.1

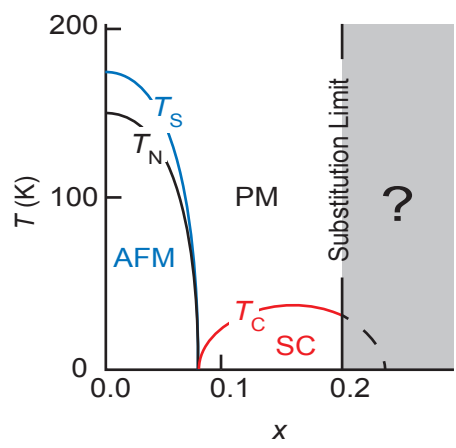

Fig.2

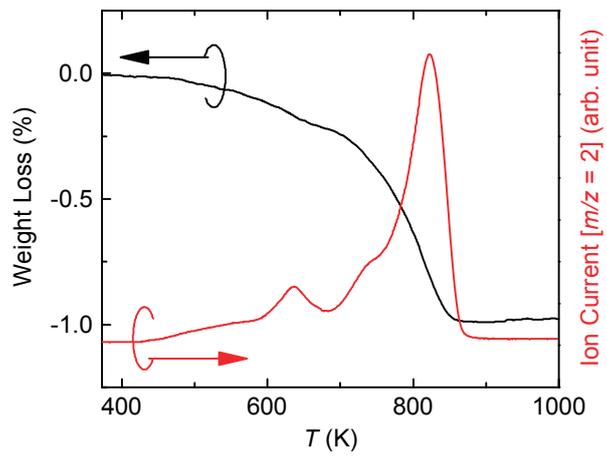

Fig.3

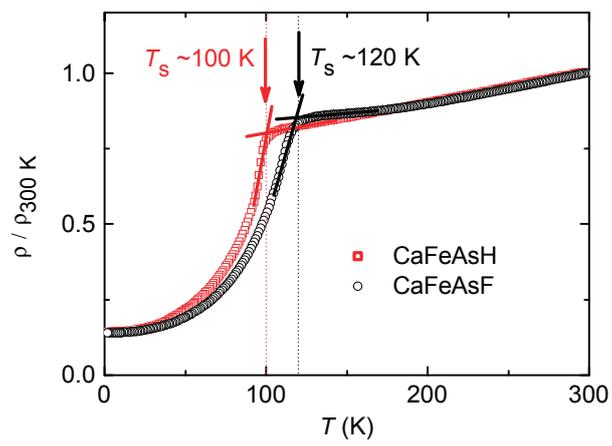

Fig.4

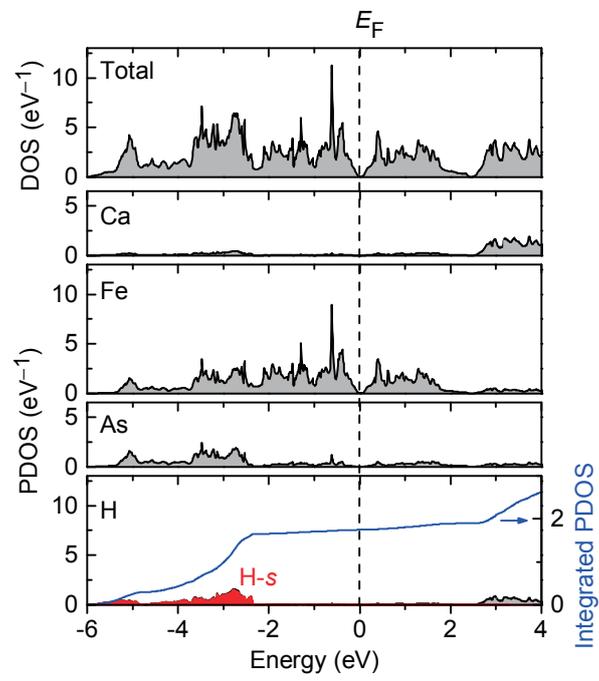

Fig.5

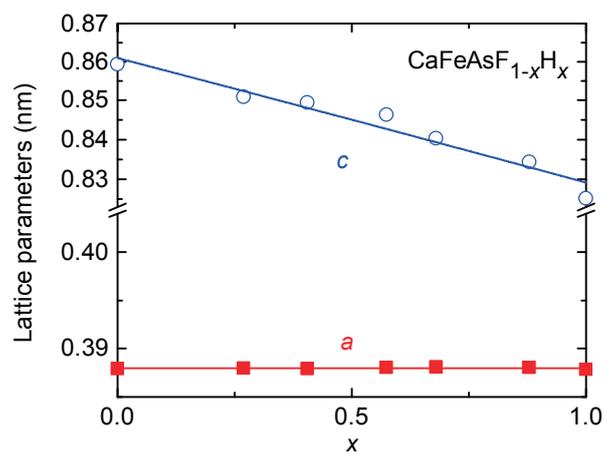

Fig.6

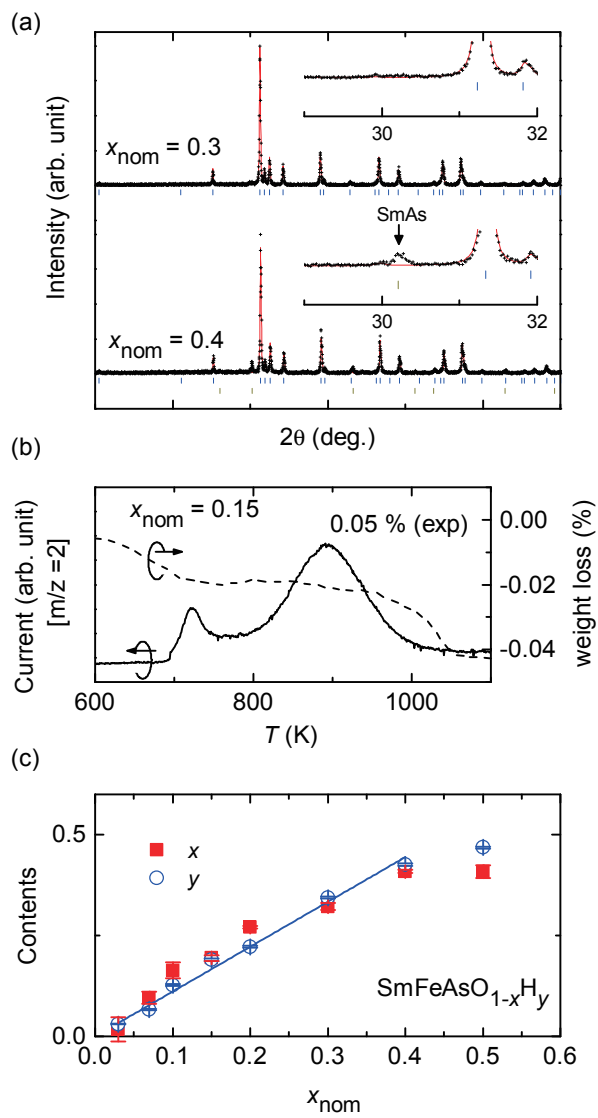

Fig.7

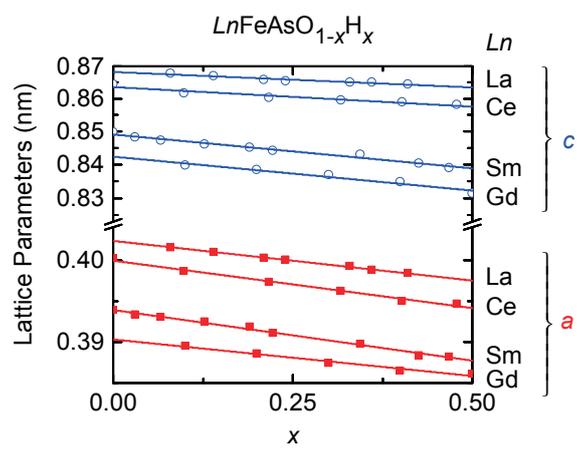

Fig.8

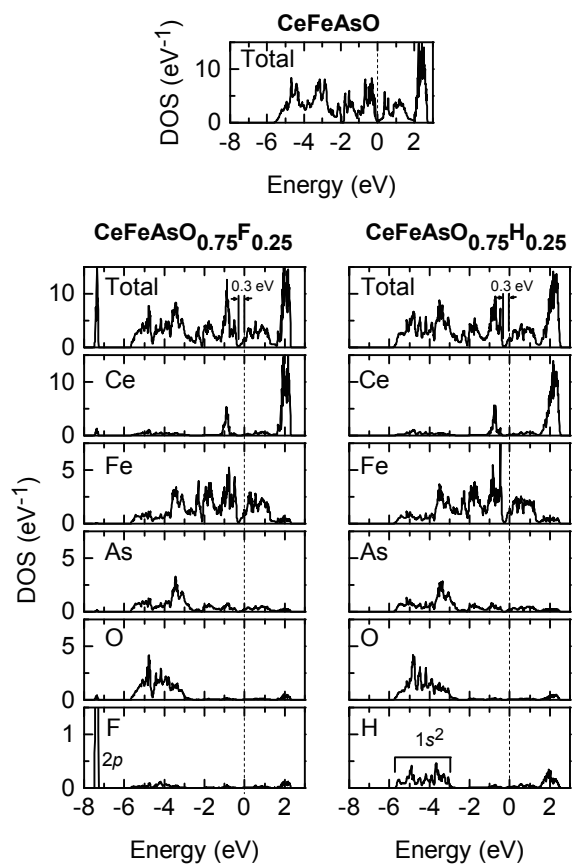

Fig.9

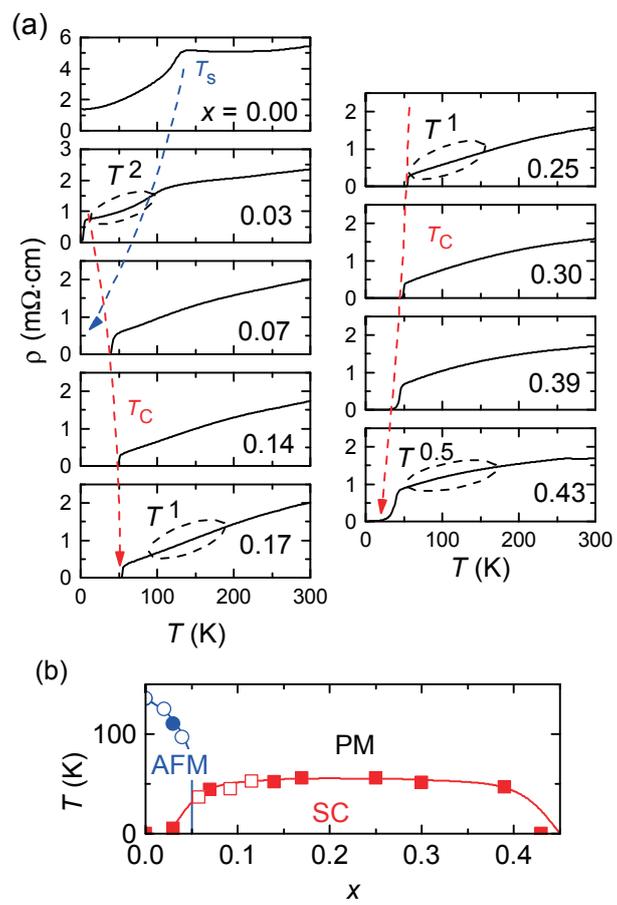

Fig.10

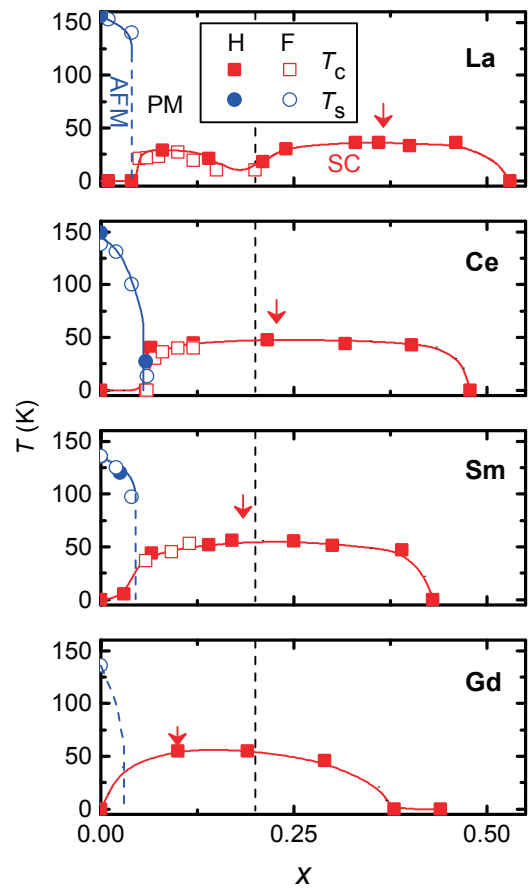

Fig.11

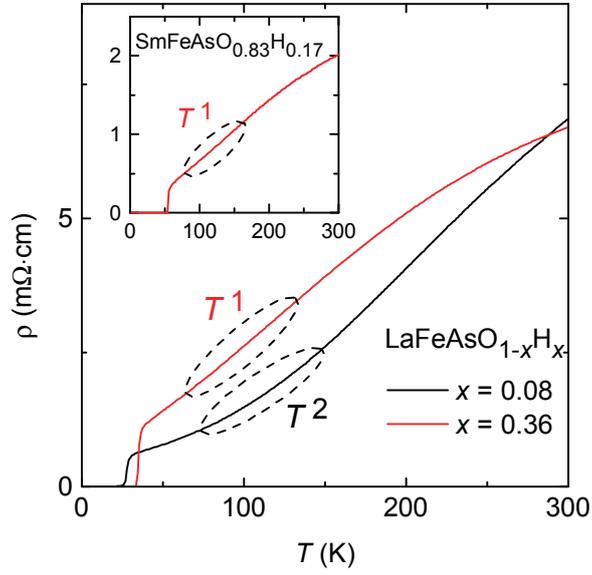

Fig.12

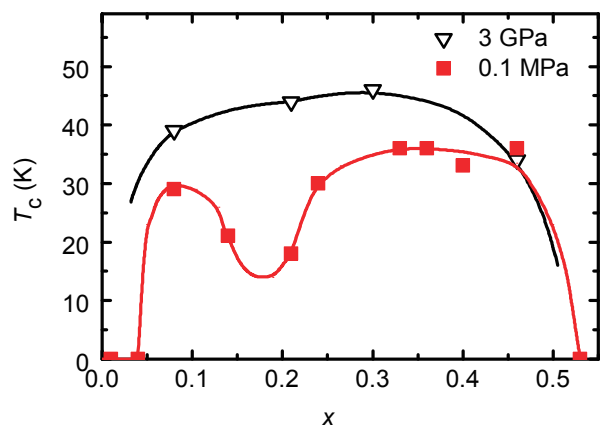

Fig.13

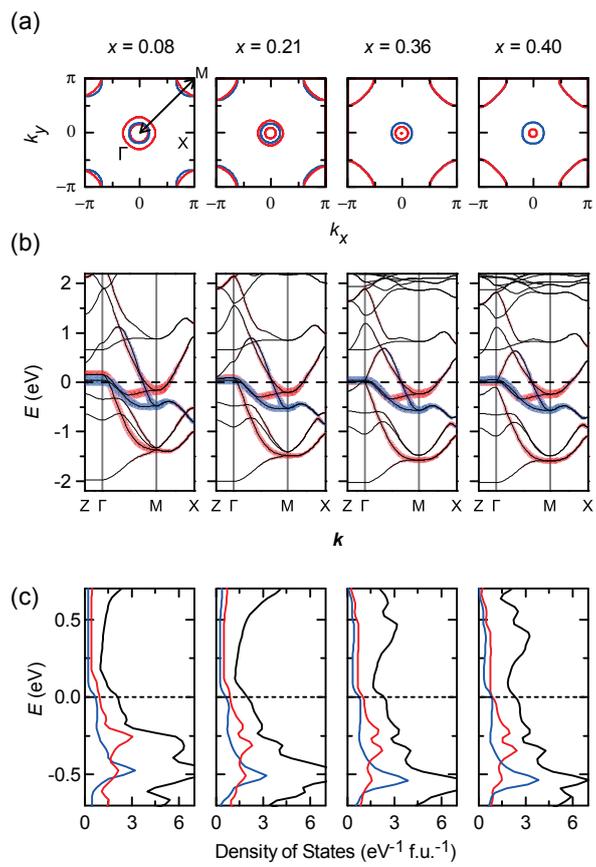

Fig.14

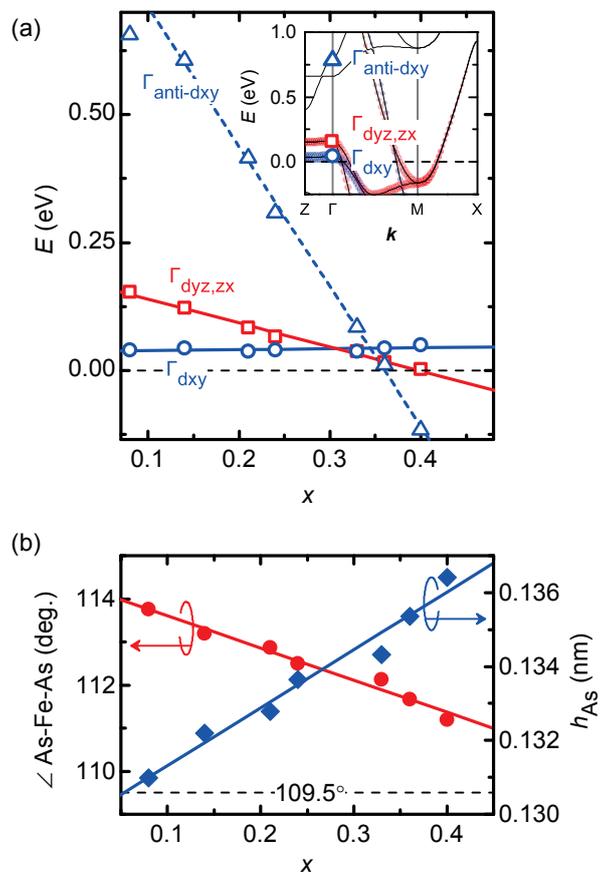

Fig.15

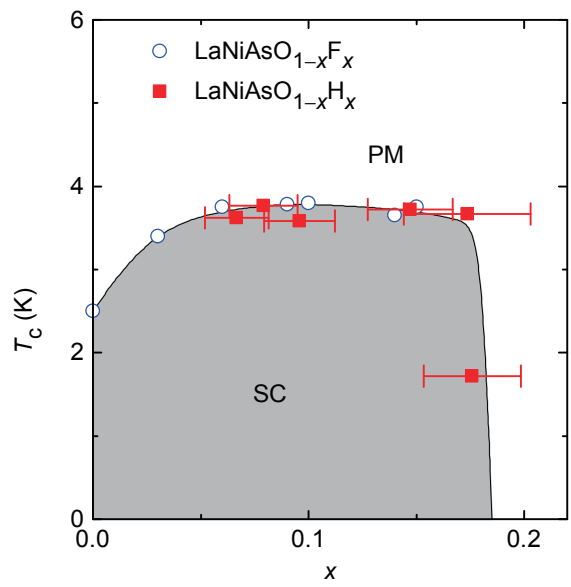

Fig.16